# Nonlinear Thermophoresis beyond Local Equilibrium Criterion

*Stefan Duhr and Dieter Braun*

*Applied Physics, Center for Nanoscience,*
*Ludwig Maximilians University München*
*Amalienstr. 54, D-80799 München, Germany*

Thermophoresis (thermodiffusion, Soret effect) moves molecules along thermal gradients. We measure its phenomenological linear drift relation by single particle tracking in convection-free settings. For moderate thermal gradients, drift velocity depends linearly on the gradient. However, for strong thermal gradients, we find a nonlinear dependence of the drift on the applied gradient for large Soret coefficient and particle radius. Interestingly, the onset of the nonlinearity coincides with a local disequilibrium of the particle. Nonlinear thermophoresis resolves several fundamental contradictions between thermophoretic experiments and theory.

*Introduction.* Nonlinear transport effects are encountered in many fields of physics. Deviations from linear response typically points towards previously unconsidered physics. In the past, thermophoretic transport, in our case the movement of particles in liquids along a temperature gradient[1-3], was treated as linear response between temperature gradient $\nabla T$ and particle drift velocity v with thermal diffusion coefficient $D_T$:

$$v = -D_T \nabla T \qquad (1)$$

We show here for the first time nonlinear transport in thermophoresis. The finding is of interest since despite ongoing experimental efforts[4-13], a generally successful theoretical explanation is lacking[14-25]. We test the linear transport of equation (1) for large particles, big Soret coefficients and steep thermal gradients by using single particle tracking. We find that for thermal gradients exceeding a distinct threshold the linearity assumption breaks down and thermophoretic drift becomes nonlinear. Interestingly, the onset of nonlinearity coincides with a transition from local thermodynamic equilibrium of the particle to local disequilibrium.

*Measurement technique.* A temperature field is generated by heating water with an infrared laser and measured microscopically with a temperature sensitive fluorescent dye[8,11,18]. We track the average velocity of single particles in a temperature gradient by fluorescence microscopy (Fig. 1a,b) and evaluate their drift velocity v against thermal gradient $\nabla T$ to test the linearity of equation (1). Imaging allows to detect particles which stick to the walls of the chamber. Thermal convection is suppressed by working in a 20 μm thin chamber, low heat conducting plastic walls (Ibidi, Munich) ensure a temperature gradient which is parallel to the chamber. The method allows to measure thermophoresis for large particles without artifacts from thermal convection.

A strong temperature gradient in a 20 µm thin water film is created optically by aqueous absorption of infrared light (Fig. 1a). A fiber coupled infrared solid state laser (Furukawa FOL1405-RTV-317, 1480 nm, 25 mW) is moderately focussed to prevent optical tweezing. Temperature is increased by 8 K in a 35 µm wide Lorentzian shape, establishing temperature gradients in the range of $\nabla T = 0...0.11$ K/µm. Heating is measured with micrometer resolution by temperature dependent fluorescence of the dye BCECF (Molecular Probes, B-1151), diluted to 50 µM in 10 mM TRIS buffer. It shows a temperature sensitivity of 2.8 %/K, consisting of 1.3 %/K pH sensitivity and thermophoresis of the dye with $S_T = 1.5\%$ K$^{-1}$. Imaging and bleaching correction have been described previously[11].

Fluorescein labelled, carboxyl modified polystyrene beads of diameter 0.5 µm and 1.9 µm (Molecular Probes, F-8827, F-8888) are desalted and diluted to picomolar concentrations to avoid particle-particle interactions. The solution is buffered in 0.5 mM TRIS at pH 7.8. At this ion concentration, particles do not stick to the chamber surfaces. Fluorescence particle images are taken at 4 Hz through a 32 x air objective (microscope Zeiss Axiotech, camera PCO SensiCam_QE) with a large depth of focus[8], allowing z-independent particle detection (Fig. 1b). Particle positions are inferred with <10 nm precision by two-dimensionally fitting the fluorescence maxima. The diffusion coefficient is inferred from unheated particle tracks and matches within 3 % the manufacturer specifications with D=0.226 µm$^2$/s (1.9 µm diameter) and D=0.858 µm$^2$/s (0.5 µm). Particles at random initial position are subjected to above-described temperature gradient which is established within <10ms. Each recording evaluated 300 single particle tracks.

To test for possible artifacts, we calculate the 3-dimensional temperature field and thermal convection flow (Fig. 2) in radial coordinates[11] using finite elements (Femlab 2.3, model file can be obtained). Simulation results are given over the experimentally used radius r = 30..150 µm. The temperature profile shows a marked horizontal and a negligible, 10-fold smaller vertical thermal gradient, leading to radial thermophoretic particle drift (Fig. 2b). Convection is sufficiently suppressed by the thin chamber, peaking at 0.05 µm/s and thus 10-fold lower than corresponding thermophoretic drift (Fig. 2c). Notably, single particle tracking cancels inward and outward convection flow to first order. Optical trapping forces were calculated[26] and yield in the worst case at the chamber center a drift of 0.01µm/s, less than 3% of the measured drift (Fig. 2d).

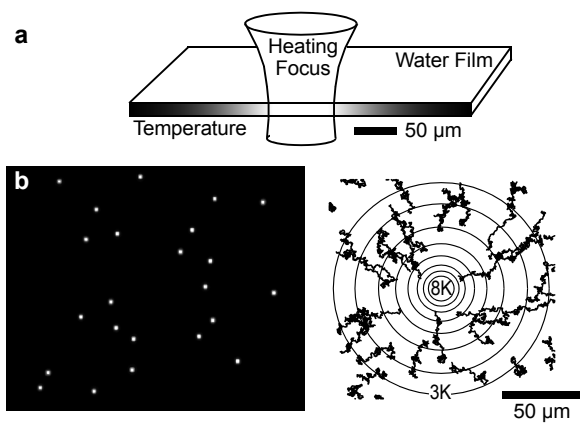

FIG. 1. *Experimental setup.* (a) A 20 µm thin water film is locally heated by infrared light. (b) The drift of single polystyrene beads is tracked in the created thermal gradient, indicated by isothermal rings (1.9 µm diameter beads in 0mM KCl).

At various radius, we collect particle drift velocities within 50 s after application of the temperature gradient (Fig. 3a,c). The time limitation ensures that the system is far from the steady state and diffusive backflow is a minor contribution: equilibration time constant is given by $b^2/D$ with an 1/e focus diameter of $b = 35$ µm leading to values of 1500 s and 5500 s for 0.5 µm and 1.9 µm particle diameters. For both particle sizes, the diffusive backflow $j = -D\nabla c$ is less than 10-times smaller than the thermodiffusive drift $j = -D_T c \nabla T$ as confirmed by numerical calculations. The steady state depletion in the heated center is estimated based on linear thermophoresis to $10^{-13}$ of bulk concentration for the small 0.5 µm diameter particles and $10^{-165}$ for 1.9 µm particles in the linear regime, rendering measurements of the steady state concentration profile prohibitive. The temperature dependence of $D_T$ of the chosen beads is rather low[18] with -2.2 %/K. To conclude, the used experimental setting allows to measure thermophoretic velocity for large single particles in strong thermal gradients without significant artifacts.

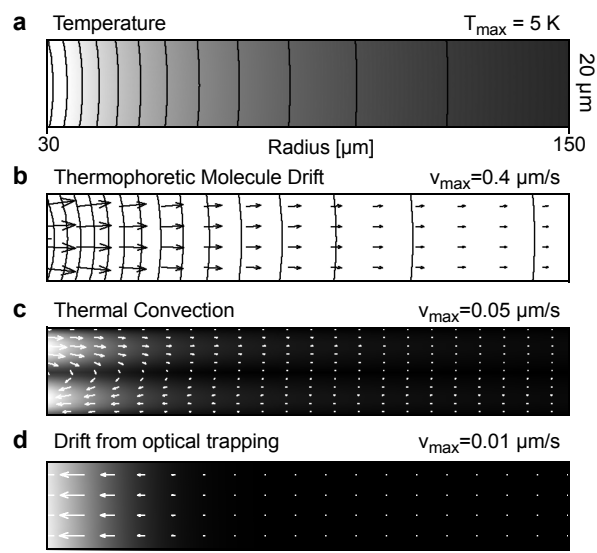

FIG. 2. *No artifacts from vertical temperature gradient or thermal convection.* (a) A 3-dimensional model of the experiment was simulated with finite element numerics. (b) The radial temperature drop dominates, leading to mostly lateral thermophoretic drift. (c) Thermal convection is 10-fold slower than thermophoretic drift. (d) Drift from optical trapping can be neglected.

*Results.* We measured the drift velocity of 0.5 µm beads in the temperature gradient and plot the results both versus experimental radius (Fig 3a) and logarithmic temperature gradient (Fig 3b). The drift velocity decreases over radius. The linear prediction of velocity v given by equation (1) with $D_T = 3.2$ µm$^2$/(sK) is plotted as solid line and fits the drift velocities very well (Fig 3a,b). This confirms the long held assumption that thermophoresis is governed by the linear approach of equation (1). However if we choose to measure larger beads with 1.9 µm diameter, the linearity breaks down. Linear scaling with $D_T = 9.8$ µm$^2$/(sK) is only found for low temperature gradients $\nabla T < 0.02$ K/µm at radius r > 100 µm (Fig. 3c). At lower radius and increased temperature gradient, the linear prediction of equation (1) is violated and beads drift with 2-fold slower than expected velocity at $\nabla T = 0.1$ K/µm (Fig. 3d).

The shape of nonlinear thermophoretic transport is extracted in Figure 4 by removing the common radius dependence and plotting versus temperature gradient $\nabla T$. To compare experiments, drift velocity v is scaled by $D_T$ of the linear regime to units of a thermal gradient K/µm. For small particles, the experiment confirms the linear drift relation of equation (1) (Fig. 4a, black), whereas for 1.9 µm particles the relation deviates nonlinearly towards slower drift velocities (Fig. 4a, white). Linear thermophoresis of 1.9 µm particles is reestablished by adding 20 mM KCl, which decreases the thermodiffusion coefficient from $D_T = 9.8$ µm²/(sK) down to $D_T = 2.2$ µm²/(sK) (Fig. 4b, black). Thus, nonlinear thermophoresis is not only a function of particle radius a, but also of thermodiffusion coefficient.

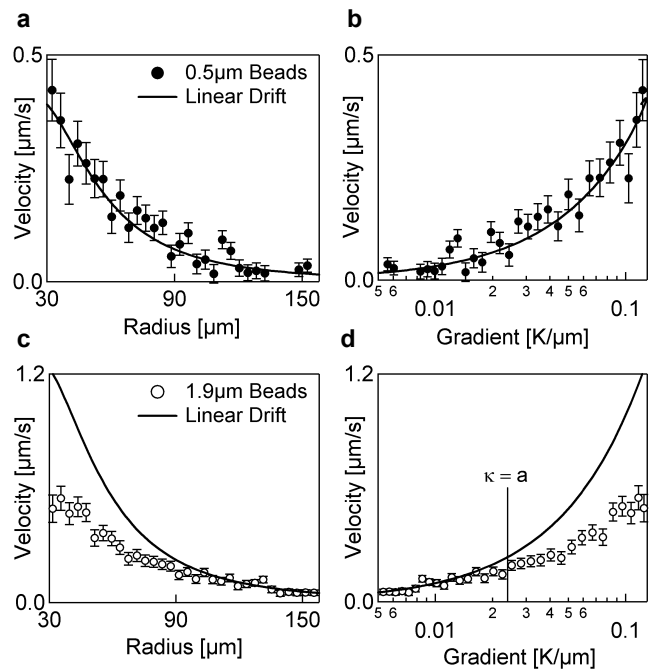

FIG. 3. *Drift velocity versus radius and temperature gradient.* (a,b) Raw data of thermophoretic drift of 0.5 µm polystyrene beads match linearly with the applied temperature gradient, both seen in the raw data over radius and a logarithmic plot against applied thermal gradients. (c,d) However, thermophoretic drift of 1.9 µm beads deviates from linear relationship for radius r = 30..100 µm where the thermal gradient exceeds 0.02K/µm.

*Discussion.* To date, the theoretical foundation of thermophoresis in liquids is under debate. We show here that the generally held assumption of linear response does not apply to thermophoresis of large particles in strong thermal gradients. We believe the findings can be coherently interpreted at three different levels of description as follows.

*Coarse graining length scale.* We start our discussion from the well documented linear gradient at low thermal gradients. As generally assumed in the past[3] and checked experimentally for shallow thermal gradients[18], the thermodiffusive steady state follows an exponential distribution

$$c/c_0 = \exp[-S_T(T - T_0)] \quad (2)$$

as the result of linear thermophoretic drift and backdiffusion. This is valid for low particle concentrations (volume ratio << 1) and temperature independent coefficients $D_T$ and D. The parameters are particle concentration c, temperature T, their boundary values $c_0$, $T_0$ and Soret coefficient $S_T = D_T/D$. The length scale κ in the form $c \propto \exp(-x/\kappa)$ of above exponential steady state is given by $\kappa = (S_T \nabla T)^{-1}$ with $S_T$ from the linear regime. We compare this depletion length with the particle radius a and find two regimes: κ » a describes a flat distribution

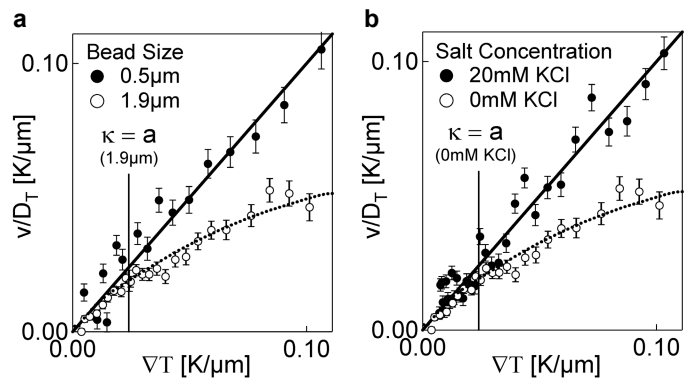

FIG. 4. *Nonlinear thermophoresis.* Normalized thermophoretic drift $v/D_T$ versus temperature gradient ∇T. (a) Beads with diameter 1.9 µm show nonlinear thermophoresis while linear transport is observed for 0.5 µm beads. (solid line: linear fit, dashed line: spline fit to guide the eye) (b) Adding 20 mM KCl switches nonlinear thermophoresis of 1.9 µm beads back to linear thermophoresis parallel to the salt-induced drop of the Soret coefficient. (solid line: linear fit, dashed line: spline fit to guide the eye)

whereas for $\kappa \ll a$ the distribution falls off steeper than the particle size itself. Interestingly, with considerable precision, the limiting case

$$\kappa \approx a \tag{3}$$

determines the temperature gradient above which thermophoretic drift becomes nonlinear. For example, for 1.9 µm particles, the limit is given by $\nabla T = 0.024 \text{K}/\mu\text{m}$ (Fig. 4a,b, horizontal line) based on the experimental values of a = 0.95 µm and $S_T$ = 43 K$^{-1}$. The value coincides with the onset of nonlinear thermophoresis. This implies that on the other side, nonlinear thermophoresis is expected for 0.5 µm particles beyond 1.1 K/µm ($S_T$ = 3.7 K$^{-1}$) and for 1.9 µm particles in 20 mM KCl beyond 0.11 K/µm ($S_T$ = 9.7 K$^{-1}$). Both limiting gradients are not experimentally accessible and linear thermophoresis is found below these values (Figure 4a,b). Thus both for variations of particle radius a and Soret coefficient $S_T$, the criteria (3) is confirmed by experiment.

We think, this finding tells us important things about the theoretical foundation of thermophoresis. While reasoning solely on coarse graining can be considered quite formal, it readily relates to discussions on the Onsager relations by van Kampen[19] and dissections of the thermophoretic steady state[18]. We provide additional physical insight with a ballistic and a thermodynamic interpretation of the onset of nonlinear thermophoresis below.

*Ballistic interpretation.* Let us compare the particle diffusion against the drift speed in the thermal gradient. In Fig. 1b, only central particles in the strong thermal gradient show an appreciable directed drift. More to the periphery, the particle's movement is dominated by diffusion and for short times, the particles nearly reversibly move back and forth along the gradient. To distinguish between ballistic and diffusive movement let us consider the time $\Delta t$ the diffusion needs to broaden the position probability distribution to the size of the particle radius: $\Delta t = a^2/D$. After $\Delta t$, linear thermophoresis moves the particle over the distance $\Delta x = v\Delta t = D_T|\nabla T|\Delta t$. If this drift is larger than the diffusive broadening $\Delta x \gg a$, we call it ballistic and if it is smaller $\Delta x \ll a$, we call it diffusive. As we see, the distinction requires again the introduction of a finite length scale, namely the radius of the particle.

The ballistic regime is arithmetically equivalent to the steep distribution limit $\kappa \ll a$ with nonlinear thermophoresis and the diffusive regime is identical to the flat distribution limit $\kappa \gg a$ with linear thermophoresis. We argue as follows. In the ballistic regime, the particles drift irreversibly along the gradient in one direction whereas in the diffusive regime, the particles move in good approximation reversibly. They fluctuate back and forth along the gradient almost like a thermally equilibrated particle. It is thus tempting to approach the linear thermophoresis regime with a local thermodynamic, "local" being defined as the length scale of the particle.

*Equilibrium interpretation.* We start by assuming that local equilibration holds in the linear thermophoresis regime. As a result, the exponential distribution (2) can be directly interpreted as Boltzmann distribution[18]

$$c/c_0 = \exp[-[G(T) - G(T_0)]/k\overline{T}] \tag{4}$$

with G(T) the Gibbs-free enthalpy of a particle in its local temperature T. However, the assumption of local equilibrium sets a limit to the gradient of G, namely that G changes less than kT over the size of the particle: $a\nabla G \ll kT$.

The equivalence of equations (2) and (4) translates such criterion to thermophoretic variables of $S_T$ and $\nabla T$. Interestingly, it sets the onset of local disequilibrium at $\kappa \ll a$. Experimentally, we thus find linear thermophoresis under conditions of local equilibrium. Nonlinear thermophoresis is found beyond local thermodynamic equilibrium. Above argument also motivates why the thermodynamically related Soret coefficient $S_T$ and not $D_T$ determines the limit equation (3).

*Towards a unified theory.* The experiments point towards combining two competing theories of thermophoresis. Below the nonlinear onset, local equilibrium theories[18,22-25] describe thermophoresis. Above the limit we propose that linear nonequilibrium theories apply which typically use fluid dynamic approaches[15-17]. Such nonequilibrium theories predict considerable slower thermal drift for the measured particles, in accordance with the nonlinear slow down of thermophoresis (Fig. 4). We think that nonlinear thermophoresis is therefore the crossover point between two linear theories: a local thermodynamic description at low thermal gradients and a nonequilibrium description at strong thermal gradients. Whether the latter is linear at even stronger temperature gradients is difficult to access by experiment.

*Scaling with particle size.* Experiments on small particles show that thermophoretic drift rises with the radius of a particle[18]. But then, large objects would move very fast in even minute thermal gradients, in strong contradiction to everyday experience. For moderate temperature gradients, the Soret coefficients of polystyrene beads were measured[18] to scale according to $S_T \propto a^2$. From equation (3) follows $\nabla T \propto a^{-3}$ and the onset of nonlinear thermophoresis becomes strongly size dependent. Macroscopic objects, even for minute gradients, then fall into the regime of nonequilibrium models where thermophoretic drift does not depend on particle size. Nonlinear thermophoresis therefore resolves the large particle paradox of local equilibrium theories.

Most techniques measure thermophoresis well below the limit given by equation (3). Only thermal field flow fractionation (TFFF)[4,6] uses exceptionally large gradients of up to 1 K/µm. In contradiction to recent measurements[18], Soret coefficients found by TFFF[20,21] scale with particle size with a variety of power laws. We see however, that the onset of nonlinear thermophoresis is highly particle size dependent ($\nabla T \propto a^{-3}$) and the nonlinearity limit is reached for larger particles by TFFF. Therefore we can expect that nonlinear thermophoresis will resolve inconsistencies between measured size scaling laws.

To conclude, we used single particle tracking to microscopically access the non-equilibrium system of particle drift in a temperature gradient. Thermophoretic drift of particles with large Soret coefficients up to $S_T$=45 K$^{-1}$ was measured. We test the linearity of thermophoretic transport and find nonlinear thermophoresis for temperature gradients above $(aS_T)^{-1}$ with a the radius of the particle. We argue that this limit marks the transition from local equilibrium[18] to local disequilibrium of particles in a thermal gradient. As a result, for the typically used tempera-ture gradients, thermophoresis should be described by local thermodynamic equilibrium[22-25] in contrast to competing theoretical approaches[15-17].

We thank Werner Köhler and Jan Dhont for discussions and Hermann Gaub for hosting our Emmy-Noether Group, which was funded by the Deutsche Forschungsgemeinschaft DFG.

1  C. Ludwig, Sitzber. Akad. Wiss. Wien, Math.-naturw. Kl. 20, 539 (1856)

2  C. Soret, Arch. Geneve 3, 48 (1879)

3  de Groot, S. R. & Mazur P. *Non-equilibrium Thermodynamics,* North-Holland, Amsterdam, 1969

4  J. C. Giddings, M. E. Hovingh and G. H. Thompson, J. Phys. Chem. 74, 4291 (1970)

5  M. Giglio and A. Vendramini, Phys. Rev. Lett. 38, 26 (1977)

6  M. E. Schimpf and J. C. Giddings, J. of Polymer Science B 27, 1317 (1989)

7  W. Köhler and P. Rossmanith, J. Phys. Chem. 99, 5838 (1995)

8  D. Braun and A. Libchaber, Phys. Rev. Lett. 89, 188103 (2002)

9  B.-J. de Gans, R. Kita, B. Müller and S. Wiegand, J. of Chem. Phys. 118, 8073 (2003)

10  J. Rauch and W. Köhler, Phys. Rev. Lett. 88, 185901 (2002)

11  S. Duhr, S. Arduini, and D. Braun, Eur. Phys. J. E 15**,** 277 (2004)

12  R. Rusconi, L.Isa, and R.Piazza, J. Opt. Soc. Am. B 21, 605 (2004)

13  S. Duhr and D. Braun, Appl. Phys. Lett. 86, 131921 (2005)

14  P. S. Epstein, Zeitschrift für Physik 54**,** 537 (1929)

15  K.I. Morozov, J. Experim. and Theor. Phys. 88, 944 (1999)

16  M. E. Schimpf and S. N. Semenov, J. Phys. Chem. B 105, 2285 (2001)

17  A. Parola and R. Piazza, Eur. Phys. J. E 15, (2004)

18  S. Duhr and D. Braun, Phys. Rev. Lett., 96, 168301 (2006)

19  N.G. van Kampen, J. Stat. Phys. 109, 471 (2002)

20  S. J. Jeon, M. E. Schimpf, and A. Nyborg, Anal. Chem. 69, 3442-3450 (1997)

21  P.M. Shiundu, G. Liu, and J.C. Giddings, Anal. Chem. 67, 2705-2713 (1995)

22  J.K.G. Dhont, J. of Chem. Phys. 120, 1632 (2004)

23  J.K.G. Dhont, J. Chem. Phys. 120, 1642 (2004)

24  Duhr and Braun, submitted

25  S. Fayolle, T. Bickel, S. Le Boiteux and A. Würger, Phys. Rev. Lett. 95, 208301 (2005)

26  Tsvi Tlusty, Amit Meller and Roy Bar-Ziv, Phys. Rev. Lett. 81, 1738 (1998)

Fig. 1.  *Experimental setup.* (a) A 20 µm thin water film is locally heated by infrared light. (b) The drift of single polystyrene beads is tracked in the created thermal gradient, indicated by isothermal rings.

Fig. 2.  *No artifacts from vertical temperature gradient or thermal convection.* (a) A 3-dimensional model of the experiment was simulated with finite element numerics. (b) The radial temperature drop dominates, leading to mostly lateral thermophoretic drift. (c) Thermal convection is 10-fold slower than thermophoretic drift. (d) Drift from optical trapping can be neglected.

Fig. 3.  *Drift velocity versus radius and temperature gradient.* (a,b) Raw data of thermophoretic drift of 0.5 µm polystyrene beads match linearly with the applied temperature gradient, both seen in the raw data over radius and a logarithmic plot against applied thermal gradients. (c,d) However, thermophoretic drift of 1.9 µm beads deviates from linear relationship for radius $r = 30..100$ µm where the thermal gradient exceeds 0.02 K/µm.

Fig. 4.  *Nonlinear thermophoresis.* Normalized thermophoretic drift $v/D_T$ versus temperature gradient $\nabla T$. (a) Beads with diameter 1.9 µm show nonlinear thermophoresis while linear transport is observed for 0.5 µm beads. (solid line: linear fit, dashed line: spline fit to guide the eye) (b) Adding 20 mM KCl switches nonlinear thermophoresis of 1.9 µm beads back to linear thermophoresis parallel to the salt-induced drop of the Soret coefficient. (solid line: linear fit, dashed line: spline fit to guide the eye)